\documentclass[superscriptaddress,twocolumn,showpacs,
preprintnumbers,amsmath,amssymb,nofootinbib,
ventriculography]{revtex4-1}
\usepackage{graphicx}

\usepackage{color}

\usepackage{amsmath}
\usepackage{amssymb}
\usepackage{amsfonts}
\usepackage[usenames,dvipsnames]{xcolor}
\usepackage{graphicx}  
\usepackage{enumerate}

\newcommand{\hh}{,\hspace{0.5cm}}

\newcommand{\eps}{\varepsilon}

\newcommand{\A}[1]{A^{\!(#1)}}                 

\newcommand{\dg}{{{n}}}                           

\newcommand{\be}{\begin{equation}}             
\newcommand{\ee}{\end{equation}}               
\newcommand{\ba}{\begin{eqnarray}}             
\newcommand{\ea}{\end{eqnarray}}               
\newcommand{\n}[1]{\label{#1}}

\begin{document}

\title{Weakly charged generalized Kerr--NUT--(A)dS spacetimes}

\author{Valeri P. Frolov}

\email{frolov@phys.ualberta.ca}

\affiliation{Theoretical Physics Institute, University of Alberta, Edmonton,
Alberta, Canada T6G 2G7}

\author{Pavel Krtou\v{s}}

\email{Pavel.Krtous@utf.mff.cuni.cz}

\affiliation{Institute of Theoretical Physics,
Faculty of Mathematics and Physics, Charles University,
V~Hole\v{s}ovi\v{c}k\'ach~2, Prague, 18000, Czech Republic}

\author{David Kubiz\v n\'ak}

\email{dkubiznak@perimeterinstitute.ca}

\affiliation{Perimeter Institute, 31 Caroline Street North, Waterloo, ON, N2L 2Y5, Canada}


\date{May 1, 2017}

\begin{abstract}
We find an explicit solution of the source free Maxwell equations in a generalized Kerr--NUT--(A)dS spacetime in all dimensions. This solution is obtained as a linear combination of the closed conformal Killing--Yano tensor $h_{ab}$, which is present in such a spacetime, and a derivative of the primary Killing vector, associated with $h_{ab}$. For the vanishing cosmological constant the obtained solution reduces to the Wald's electromagnetic field generated from the primary Killing vector.
\end{abstract}


\maketitle


Electromagnetic fields in the vicinity of (rotating) black holes in four dimensions have interesting astrophysical applications and
have been investigated by many authors, see e.g. \cite{Wald:1974np,King:1975tt, bicak1977stationary, bivcak1976stationary, bivcak1980stationary, bivcak1985magnetic, Aliev:1989wx, Penna:2014aza}. See also \cite{Aliev:2004ec, Ortaggio:2004kr, Aliev:2006tt,Ortaggio:2006ng,  Aliev:2007qi, Chen:2007fs, Krtous:2007} for the studies in higher dimensions.

Let us first recall the Wald construction of a weak electromagnetic field in spacetimes with symmetry \cite{Wald:1974np}. Consider a test electromagnetic field in a $D$-dimensional curved spacetime with metric $g_{ab}$. The corresponding source-free Maxwell equations are
\be\n{FF}
F^{ab}_{\ \ ;b}=0\hh F_{[ab;c]}=0\, .
\ee
The latter equation implies that the electromagnetic field strength can be found as an external derivative of the potential 1-form $A_a$, $F_{ab}=A_{b,a}-A_{a,b}$. Imposing the Lorenz gauge condition one can write the first Maxwell equation in the form
\be\n{AA}
 A_{a}^{\ ;a}=0\hh\Box A_a-R_{a}^{\ c} A_c=0\, ,
\ee
where  $\Box A_a=\nabla_b \nabla^b A_a$.

Suppose now that the spacetime possesses a Killing vector $\xi^a$. Then the Killing equation $\xi_{(a;b)}=0$ and its integrability conditions imply
\be\n{XX}
\xi_a^{\ ;a}=0\hh\Box \xi_a+R_{a}^{\ c} \xi_c=0\, .
\ee
The equations \eqref{AA} and \eqref{XX} are quite similar. The only difference is the sign of the curvature term.
This means that any Killing vector in a vacuum spacetime serves as a
vector potential for a test source-free Maxwell field. Therefore a special set of test source-free electromagnetic fields in the background of vacuum spacetimes can be generated by simply using the isometries of these spacetimes.
In such a way one can generate a weakly charged Kerr black hole, or immerse this black hole in a `uniform magnetic field' \cite{Papapetrou:1966zz,Wald:1974np}.

In this letter we show how to generalize the Wald approach to a wide class of spacetimes with a cosmological constant  that admit a closed conformal Killing--Yano (CCKY) tensor $h_{ab}$  of rank 2. Such spacetimes obey the Einstein equation
\be\n{RL}
R_{ab}={2\over D-2} \Lambda g_{ab}\,
\ee
with the cosmological constant $\Lambda$, and admit a CCKY 2-form $h_{ab}$ satisfying the equation
\be\n{hh}
h_{ab;c}=g_{ca}\xi_b-g_{cb}\xi_a\, ,
\ee
where
\be\n{hX}
\xi_a={1\over D-1}h^b_{\ a;b}\, .
\ee

Since $h_{ab}$ is closed, $h_{[ab,c]}=0$, there exists a potential one form $b_a$, such that
\be
h_{ab}=b_{b,a}-b_{a,b}\, .
\ee
Using the integrability condition for \eqref{hh} one obtains
\be\n{XR}
\xi_{(a;b)}={1\over 2(D-2)}\left(
h_{ac} R^c_{\ b}+h_{ab} R^c_{\ a}
\right)\, .
\ee
Thus, in an Einstein space, that is when \eqref{RL} is satisfied, it follows that
\begin{equation}\label{KillEq}
\xi_{(a;b)}=0\,,
\end{equation}%
and hence $\xi^a$ is a Killing vector. To distinguish it from other Killing vectors which may be present in a given spacetime, we call it a \emph{primary} Killing vector.

Let us now define the following 2-form constructed from $h_{ab}$ and the primary Killing vector $\xi^a$:
\be\n{ff}
{\cal F}_{ab}=2q(\xi_{[b,a]}+\lambda h_{ab})\hh \lambda={2\Lambda\over (D-1)(D-2)}\,,
\ee
where the constant $q$ parameterizes the field strength.
This 2-form is closed and can be generated from a 1-form potential
\be\n{q}
{\cal A}_a=q(\xi_a + 2\lambda b_a)\, .
\ee
Moreover, using relations \eqref{XX},\eqref{RL}, and \eqref{hX} one can easily check that ${\cal F}_{ab}$ is also divergence free,
\be\n{fM}
{\cal F}^{ab}_{\ \ ;b}=0\, .
\ee
In other words, in any Einstein space with the  CCKY tensor $h_{ab}$ there always exists a solution of the source-free Maxwell equations
that can be written in the form~\eqref{ff}. Since this electromagnetic field is generated from the primary Killing vector, we call it a {\em primary} electromagnetic field.


An important special case occurs when $h_{ab}$ is non-degenerate. Let us denote $D=2n+\eps$ the number of spacetime dimensions, with $\eps=0$ in even dimensions and $\eps=1$ in odd dimensions. Then the non-degeneracy condition means that the tensor $Q^a{}_b=h^{ac} h_{bc}$  has $n$ independent nonvanishing {eigenvalues $x_{\mu}^2$} $(\mu=1,\ldots,n)$. We call such a CCKY tensor a \emph{principal tensor}. The most general solution of the Einstein equations \eqref{RL} which possesses the principal tensor is the so called {\em Kerr--NUT--(A)dS metric} \cite{Houri:2007uq, Houri:2007xz, Krtous:2008tb, Yasui:2008zz, Houri:2008ng}. This solution was found in \cite{Chen:2006xh}. It contains the following set of `free' parameters: mass $M$, $(n-1+\eps)$ rotation parameters, $a_i$, and {$(n-1)$ NUT parameters, $N_\mu$}. For $M=N_\mu=0$ the Kerr--NUT--(A)dS metric becomes the metric of an (anti)de Sitter spacetime, and when also $\Lambda=0$ it is a flat metric.

The Kerr--NUT--(A)dS spacetime has a number of remarkable properties. In particular, besides the primary Killing vector $\xi^a$, it has $n-1+\eps$ additional commuting Killing vectors. A set of $n$ equations $x_{\mu}$=const defines a $(n-1+\eps)$-dimensional submanifold, such that all Killing vectors are tangent to it. If one introduce coordinates $\psi_i$ on this submanifold by conditions that the Killing vectors take the form $\partial_{\psi_i}$, then $(x_{\mu},\psi_i)$ can be used as spacetime coordinates, which are called \emph{canonical}. A remarkable property of the Kerr--NUT--(A)dS metric is that when written in canonical coordinates the tensor components of $h_{ab}$ do not depend on the mass and NUT parameters. That is, they are the same as the corresponding components of this tensor in the (A)dS or flat spacetime. Thus, the operation \eqref{ff} of `upgrading' the Killing vector ansatz for an electromagnetic field can be interpreted as a subtraction from $2\xi_{[b,a]}$ a similar quantity, calculated for the corresponding (anti-)de Sitter background metric. Namely this prescription was used in \cite{Aliev:2006tt, Aliev:2007qi} for obtaining the weakly charged Kerr--(A)dS black holes in all dimensions.

In order to obtain the components of the field ${\cal F}_{ab}$ in the canonical coordinates we introduce a non-normalized frame of 1-forms \cite{Krtous:2007}
\begin{equation}\label{frame}
    \epsilon^\mu = d x_\mu \,,\quad\!
    \epsilon^{\hat\mu} = \sum_{k=0}^{n-1} \A{k}_\mu d\psi_k\,,\quad\!
    \epsilon^{\hat0} = \sum_{k=0}^{n} \A{k} d\psi_k\,.
\end{equation}
The last 1-form $\epsilon^{\hat0}$ is present only in odd dimensions. The functions $\A{k}$ and $\A{k}_\mu$ are symmetric polynomials in $x_\mu^2$,
\begin{equation*}
  \A{k}=\!\!\!\!\!\sum_{\substack{\nu_1,\dots,\nu_k=1\\\nu_1<\dots<\nu_k}}^\dg\!\!\!\!\!x^2_{\nu_1}\dots x^2_{\nu_k}\,,
\quad
  \A{j}_{\mu}=\!\!\!\!\!\sum_{\substack{\nu_1,\dots,\nu_j=1\\\nu_1<\dots<\nu_j\\\nu_i\ne\mu}}^\dg\!\!\!\!\!x^2_{\nu_1}\dots x^2_{\nu_j}\,.
\end{equation*}

In this frame $h_{ab}$ takes the form \cite{Kubiznak:2006kt}
\begin{equation}
h=\sum_{\mu=1}^n x_{\mu} \epsilon^\mu \wedge \epsilon^{\hat\mu}\, .
\end{equation}
It is possible to show that the strength tensor ${\cal F}_{ab}$ has a similar form
\be
{\cal F}=\sum_{\mu=1}^n f_{\mu} \epsilon^\mu \wedge \epsilon^{\hat\mu}\, ,
\ee
where $f_{\mu}$ are functions of all coordinates $x_{\nu}$.
Hence this field is a special case of a wide class of solutions of the Maxwell equations which are `aligned with' the CCKY tensor $h_{ab}$. This class was described and studied in details in \cite{Krtous:2007}. Using the results of this paper one obtains
\be\label{fmu}
f_{\mu}=\phi_{,\mu}\hh \phi= -2q\sum_{\mu=1}^n \frac{N_\mu x_\mu^{1-\eps}}{U_\mu}\, .
\end{equation}
Here, $N_\mu$ are NUT and mass parameters\footnote{%
Here, we set $N_n=i M$. It corresponds to the convention, that the radial coordinate is $x_n=ir$. See \cite{Houri:2007uq, Houri:2007xz, Krtous:2008tb, Yasui:2008zz, Houri:2008ng}.}
of the Kerr--NUT--(A)dS solution and functions $U_\mu$ are just polynomials in $x_\mu^2$
\begin{equation}\label{Udef}
    U_{\mu}=\prod_{\substack{\nu=1\\\nu\ne\mu}}^n(x_{\nu}^2-x_{\mu}^2)\,.
\end{equation}
Up to pure gauge terms, the corresponding vector potential \eqref{q} reads
\begin{equation}\label{vectpot}
    \mathcal{A} = -2q\sum_{\mu=1}^n \frac{N_\mu x_\mu^{1-\eps}}{U_\mu}\epsilon^{\hat\mu}\, .
\end{equation}
In four dimensions it is possible to `backreact' this electromagnetic field on the geometry, by simply modifying the metric functions, to obtain a fully charged Kerr--NUT--(A)dS spacetime \cite{Carter:1968pl,plebanski1975class}. In higher dimensions, though, this is no longer possible within the realms of pure Einstein--Maxwell theory, see however \cite{Chong:2005hr}, and additional fields
have to be introduced, e.g. \cite{Chow:2008fe}.

If one omits the restriction that the CCKY tensor $h_{ab}$ is non-degenerate, the class of the
metrics which possess such a tensor and obey the Einstein equations \eqref{RL}
becomes much larger. These solutions, called \emph{generalized Kerr--NUT--(A)dS} metrics, were described in \cite{Houri:2008th, Houri:2008ng, Oota:2008uj, Yasui:2011pr}.
They describe a huge family of geometries, ranging from the K\"ahler metrics, Einstein--Sasaki geometries, generalized Taub--NUT metrics, or rotating black holes with some equal and some vanishing rotation parameters. As evident from the discussion, our construction of the solution \eqref{ff} for the test electromagnetic field works in these metrics as well.

\section*{Acknowledgments}

V.F.\ thanks the Natural Sciences and Engineering Research Council of Canada and the Killam Trust for financial support.
P.K.\ is supported by the project of excellence of the Czech Science Foundation \mbox{No.~14-37086G}.
D.K.\ is supported by the Perimeter Institute for Theoretical Physics and by the Natural Sciences and Engineering Research Council of Canada. Research at Perimeter Institute is supported by the Government of Canada through
the Department of Innovation, Science and Economic Development Canada and by
the Province of Ontario through the Ministry of Research, Innovation and Science.

\providecommand{\href}[2]{#2}\begingroup\raggedright\endgroup

\end{document}